\documentclass[12pt]{article}
\usepackage{amsmath,amssymb,bm,epsf,epsfig,graphicx}
\input epsf.sty
\usepackage[utf8]{inputenc}
\usepackage[english]{babel}
\usepackage{amsmath}
\usepackage{latexsym} 
\usepackage{amsfonts}
\usepackage{mathtools}
\usepackage{amssymb}
\usepackage[left=3cm,right=3cm,top=3.1cm,bottom=3.1cm]{geometry}
\usepackage{cite}

\usepackage{hyperref}
\numberwithin{equation}{section}

\newcommand{\Tr}{\mathop{\rm Tr}\nolimits}

\newcommand{\ad}{\mathop{\rm ad}\nolimits}

\def\bra#1{\langle #1 |}
\def\ket#1{|#1 \rangle}

\def\Trr#1{\Tr\left[\, #1 \,\right]}

\def \be {\begin{eqnarray}}
\def \ee {\end{eqnarray}}
\def \bdm {\begin{displaymath}}
\def \edm {\end{displaymath}}

\def \tr{{\rm tr}}

\def\del {\partial}
\def\0{\nonumber}

\def \VV {{\mathbb V}}

\def \HH {{\mathbb H}}
\def\bj {\bar \j}
\def\bi {\bar \i}
\def\bm{\bar m}

\begin{document}
{}~ \hfill\hbox{ARC-18-02} \break

\vspace*{3.1cm}

\centerline{\Large \bf Localization of effective actions}\vspace{.4cm}
\centerline{\Large \bf in open superstring field theory} \vspace{.2cm}
\vspace*{.1cm}

\begin{center}

{\large Carlo Maccaferri\footnote{Email: maccafer at gmail.com} and Alberto Merlano\footnote{Email: albemerlano at gmail.com} } 
\vskip 1 cm
{\it Dipartimento di Fisica, Universit\'a di Torino, \\INFN  Sezione di Torino and Arnold-Regge Center\\
Via Pietro Giuria 1, I-10125 Torino, Italy}
\end{center}

\vspace*{6.0ex}

\centerline{\bf Abstract}
\bigskip
We consider the construction of  the algebraic part of D-branes tree-level effective action from Berkovits open superstring field theory. Applying this construction to the quartic potential of massless fields carrying a specific worldsheet charge, we  show that the full contribution to the potential  localizes at the boundary of moduli space, reducing to elementary two-point functions. As examples of this general mechanism, we show how  the Yang-Mills quartic potential and the instanton effective action of a $Dp/D(p-4)$ system are reproduced. \vfill \eject

\baselineskip=16pt

\tableofcontents
\newpage
\section{Introduction}\label{intro}
In recent years, either starting from a small-Hilbert space \cite{Witten-super} or  large Hilbert space  \cite{Berkovits} approach, there has been a lot of progress in the formulation of superstring field theories 
\cite{ Erler:2017onq, Ohmori:2017wtx, Kunitomo:2016kwh, Erler:2016rxg, Konopka:2016grr, Erler:2016ybs, Sen:2015uaa, Kunitomo:2015usa, EKS, ONT
}.

In this paper we focus on the construction of tree-level effective actions for the massless states of the open superstring.  Our starting point will be the WZW-like action by Berkovits \cite{Berkovits}, in the large Hilbert space. In this regards, an analytic construction of the Yang-Mills quartic potential  has been given some time ago by Berkovits and Schnabl \cite{BS}\footnote{This computation has been very recently generalized to the Ramond sector in \cite{kishimoto}, using the complete action of \cite{OK}}. 

Focussing on the algebraic couplings between the massless fields up to quartic order, we find out that
their computation is fully captured by singular terms at the boundary of moduli space.
More precisely, when we are computing the tree-level effective potential of a massless string field $\Phi_A$  which can be decomposed in eigenstates of a world-sheet charge $J_0$ with eigenvalues $\pm1$ (an assumption which includes  the non-abelian gauge field and also many other interesting cases such as the instanton moduli on a $Dp$-$D(p-4)$ system, or D-branes moduli on a Calabi-Yau)) 
\be
&&\Phi_A=\Phi_A^{(+)}+\Phi_A^{(-)},\\
&&J_0\Phi_A^{(\pm)}=\pm\, \Phi_A^{(\pm)},
\ee
we find out that, because of the failure of the propagator to truly invert the BRST charge, the whole quartic potential localizes at the boundary of moduli space. If not for this singular term, the amplitude would identically vanish. Concretely, we  show that the full effective potential at quartic order reduces to simple two-point functions
\be
S_{\rm eff}^{(4)}(\Phi_A)=\frac18\left[\left\langle\widehat h^{(--)}{\Big |}h^{(++)}\right\rangle+\left\langle\widehat g^{(+-)}{\Big |}g^{(-+)}\right\rangle+{\Big (}(+)\leftrightarrow(-){\Big)}\right].
\ee
The  ($h$, $g$) fields  are the projection to level zero of the following star products
\be
h^{(\pm\pm)}&=&P_0[\Phi_A^{(\pm)},Q_B\Phi_A^{(\pm)}],\quad J_0=\pm2,\\
\widehat h^{(\pm\pm)}&=&P_0[\Phi_A^{(\pm)},\eta_0\Phi_A^{(\pm)}],\quad J_0=\pm2,\\
g^{(\pm\mp)}&=&P_0[\eta_0\Phi_A^{(\pm)},Q_B\Phi_A^{(\mp)}],\quad J_0=0,\\
\widehat g^{(\pm\mp)}&=&P_0[\Phi_A^{(\pm)},\Phi_A^{(\mp)}],\quad J_0=0,
\ee
 where $P_0$ is the projector on the kernel of $L_0$ and $[\ ,\ ]$ is the graded commutator with respect to Witten star product. 
 
This mechanism is interesting {\it per se},  but a very practical  advantage  is that the $(h,g)$ fields can be readily computed by  leading order OPE and one doesn't need to know the full four-point function of the massless field $\Phi_A$ to compute its quartic potential. \\

The paper is organised as follows. In section \ref{sec2} we set up the construction of the tree-level effective action in Berkovits WZW-like open superstring field theory and we focus on the quartic potential of the massless fields. We show that picture changing at zero momentum is subtle and in general develops singular contributions at the boundary of moduli space, which can in fact account for the full effective action. In section \ref{sec3} we give a proof that the effective action is  indeed localized at the boundary, when the string field entering the quartic potential can be expressed as the sum of two fields of opposite $R$ charge in the ${\cal N}=2$ description of the string background under consideration. Section \ref{sec4} contains two non-trivial examples of this localization mechanism: the quartic Yang-Mills potential and the quartic action for the instanton-moduli of a $D3$--$D(-1)$ system. We  conclude in section \ref{concl} with some comments and future directions. Appendix \ref{appA} contains the detailed computation of the localization of the action and Appendix \ref{appB} is a collection of our conventions and of useful formulas needed in the main text.

\section{Tree level effective action}\label{sec2}

We start with the classical action for Neveu-Schwarz  open strings on a given $D$-brane system in the large Hilbert space \cite{Berkovits, BOZ}
\begin{equation}
S[\Phi]=-\int_{0}^{1}dt\, \Tr[(\eta_0A_t)A_Q]\label{action},
\end{equation}
in terms of the connections
\begin{equation}
A_t(t) = e^{-t\Phi}\partial_t e^{t\Phi},
\end{equation}
\begin{equation}
A_Q(t) = e^{-t\Phi}Q_Be^{t\Phi}.
\end{equation}
Here $Q_B=\oint\frac{dz}{2\pi i} J_{\rm BRST}(z)$ is the BRST charge of the RNS superstring, and $\eta_0=\oint\frac{dz}{2\pi i}\eta(z)$ is the zero mode of the $\eta$--ghost, \cite{FMS}.  $\Phi$ is the dynamical open string field which is a generic ghost and picture number zero state in the Large Hilbert Space
\be
\eta_0\Phi\neq0,\quad\quad{\rm generically}.
\ee
$\Tr$ is Witten integration \cite{Witten} in the large Hilbert space and $*$ product is always understood when string-fields are multiplied. Finally the BRST charge $Q_B$ and $\eta_0$ are mutually anti-commuting graded derivations with respect to the star product and the Witten integration of $Q_B \,/ \eta_0-$exact terms is identically zero.  

Since we will follow a perturbative approach, it is convenient to expand the action (\ref{action}) in ``powers'' of the dynamical  string field $\Phi$.
To do so we observe that, given a star algebra derivative $D$, its $t$-dependent connection $A_D(t)$ can be explicitly written as
\be
A_D(t)&=& e^{-t\Phi}De^{t\Phi}=e^{-t\Phi}\int_0^t ds \ e^{s\Phi} (D\Phi)   e^{(t-s)\Phi}= \int_0^tds\  e^{-(t-s)\ad_\Phi}(D\Phi)\0\\
&=& \int_0^tds\  e^{-s\ad_\Phi}(D\Phi)=  \frac{1-e^{-t\ad_\Phi}}{\ad_\Phi}(D\Phi)\0\\
&=&  \sum \limits_{n=1}^{+\infty} \dfrac{(-1)^{n-1}t^n}{n!} \ad_\Phi^{\,n-1}(D\Phi),
\ee
where the adjoint action is defined through a graded $*$-commutator
\be
\ad_\phi(\chi)\equiv [\phi,\chi]=\phi*\chi-(-1)^{|\phi|\,|\chi|}\chi*\phi.
\ee
Using this representation for the two connections $A_t$ and $A_Q$ and performing the $dt$ integral in \eqref{action} we end up with
\be
S[\Phi]=-\frac12\Tr[(\eta_0\Phi)(Q_B\Phi)]-\sum_{n=1}^{+\infty}\dfrac{(-1)^n} {(n+2)!} \Tr[(\eta_0\Phi)\ad^{\,n}_{\Phi}\,(Q_B\Phi)],
\ee
where we have isolated the kinetic term from the infinite tower of interaction vertices.
We are interested in computing the tree-level effective potential for the zero momentum part of a massless field. To do so we 
fix the standard  gauge
\be
b_0\Phi&=&0,\\
\xi_0\Phi&=&0,
\ee
which means
\be
\Phi=\xi_0\Psi,
\ee
with
\be
b_0\Psi&=&0\\
\eta_0\Psi&=&0\\
pict[\Psi]&=&-1.
\ee
A prototype example of the kind of massless field we are interested in is the zero momentum part of the gauge field living on the world-volume of a $Dp$-brane, which in our setting is represented by the string field
\be
\Phi_A=A_\mu \xi  \, c \, \psi^\mu e^{-\phi}=A_\mu c\, \gamma^{-1}\psi^\mu,\label{gaugefield}
\ee
where $A_\mu$ are constant Chan-Paton matrices in $U(N)$. Notice that $\psi^\mu$ is a superconformal matter primary of weigth $1/2$ and it is in fact part of a worldsheet $\mathcal{N}=1$ superfield $\psi^\mu+\theta(i\sqrt2 \del X^\mu)$. 
The string field $\Phi_A$ is on-shell  in the large Hilbert space
\be
\eta_0 Q_B\Phi_A=0.
\ee
Notice that\footnote{Throughout this paper we set $\alpha'=1$.}
\be
\eta_0\Phi_A&=&A_\mu \, c\, \psi^\mu e^{-\phi}=A_\mu\, c \, \psi^\mu\delta(\gamma),\\
Q_B\Phi_A&=&A_\mu\left(c(i\sqrt2 \,\del X^\mu)-\gamma\, \psi^\mu\right),
\ee
are the physical vertex operators at picture $-1$ and $0$ respectively, in the small Hilbert space.

This choice of $\Phi_A$ can be generalized to a string field of the form
\be
\Phi_A=c\gamma^{-1} \VV_{\frac12},\label{stringfield}
\ee
where $\VV_{\frac12}$ is a grassmann-odd superconformal {\it matter} primary of weight $1/2$
\be
T(z)\VV_{\frac12}(0)=\frac{\frac12 \VV_{\frac12}(0)}{z^2}+\frac{\del\VV_{\frac12}(0)}{z}+\textrm{regular}.
\ee
 This means that there exists a grassmann-even  world-sheet superpartner $\VV_1$ of weight 1 such that
\be
T_F(z)\VV_{\frac12}(0)=\frac{\VV_1(0)}{z}+\textrm{regular},
\ee
where $T_F(z)$ is the supercurrent of the matter $\mathcal{N}=1$ SCFT. This generic construction implies the physical condition
\be
\eta_0 Q_B\Phi_A=0
\ee
and
\be
\eta_0\Phi_A&=&c\VV_{\frac12}e^{-\phi}\label{stringfieldeta}\\
Q_B\Phi_A&=&c\VV_{1}-\gamma\VV_{\frac12},\label{stringfieldQ}
\ee
just as in the case of the zero-momentum gauge field. Notice in addition that $\Phi_A$ is in the kernel of the total matter+ghosts $L_0$
\be
L_0\Phi_A=0.
\ee
\subsection{Integrating out the massive fields}
Our aim is to obtain an effective action for the spacetime fields which are encoded in the worldsheet field $\VV_{\frac12}$. To this end, following the notation by Berkovits and Schnabl \cite{BS}, we split the string field as
\be
\Phi=\Phi_A+R,
\ee
where $R$ contains the massive fields.
This decomposition can be performed using the projector on the kernel of $L_0$ which we denote $P_0$
\be
\Phi_A&=&P_0\Phi,\\
R&=&(1-P_0)\Phi\equiv\bar P_0\Phi,
\ee
and we obviously have
\be
P_0+\bar P_0&=&1,\\
P_0\bar P_0&=&0.
\ee
As in standard  field theory, the tree-level effective action for $\Phi_A$ is  given by solving the equation of motion for $R$ in the form $R=R(\Phi_A)$ and then computing the 
classical action 
\be
S^{\rm tree}_{\rm eff}[\Phi_A]=S_{\rm class}[\Phi_A+R(\Phi_A)].
\ee
To obtain the equation of motion for $R$ consider the  variation of the action, which we schematically write as
\be
\delta S&=&\Tr[\delta\Phi\ \textrm{EOM}(\Phi)]=\Tr[\delta\Phi (P_0+\bar P_0)\ \textrm{EOM}(\Phi)]\\
&=&\Tr[(\delta\Phi_A+\delta R)\ \textrm{EOM}(\Phi)],
\ee
which implies that the $R$-equation is the projection outside the kernel of $L_0$ of the full equation of motion
\be
\frac{\delta S}{\delta R}=\bar P_0\, \textrm{EOM}(\Phi_A+R)=0.
\ee
The explicit form of the string field ``$\textrm{EOM}$'' depends on the  action \eqref{action} and  for our purposes it can be written as
\be
\textrm{EOM}(\Phi)=\frac{e^{\ad_\Phi}-1}{\ad_\Phi}\ \eta_0(e^{-\Phi}Q_Be^{\Phi})= \frac{e^{\ad_\Phi}-1}{\ad_\Phi}\ \eta_0\ \frac{1- e^{-\ad_\Phi}}{\ad_\Phi}\ Q_B\Phi.
\ee
The operators appearing above  are perturbatively defined as
\be
\frac{e^{\ad_\Phi}-1}{\ad_\Phi}&=&1+\frac12\ad_\Phi+\frac16\ad_\Phi^2+\frac1{24}\ad_\Phi^3+\cdots\\
\frac{1-e^{-\ad_\Phi}}{\ad_\Phi}&=&1-\frac12\ad_\Phi+\frac16\ad_\Phi^2-\frac1{24}\ad_\Phi^3+\cdots\ .
\ee
 Notice that, because $\frac{e^{\ad_\Phi}-1}{\ad_\Phi}$ is an invertible operator on the star-algebra, we have that
\be
\textrm{EOM}(\Phi)=0\quad\leftrightarrow\quad\eta_0(e^{-\Phi}Q_B e^{\Phi})=0,
\ee 
which is the well known equation of motion of the Berkovits OSFT. However the prefactor $\frac{e^{\ad_\Phi}-1}{\ad_\Phi}$ cannot be ignored when we are interested in the projection outside the kernel of $L_0$, and the equation for $R$ thus reads
\be
\bar P_0\  \frac{e^{\ad_\Phi}-1}{\ad_\Phi}\ \eta_0\ \frac{1-e^{-\ad_\Phi}}{\ad_\Phi}\ Q_B \Phi{\Bigg |}_{\Phi=\Phi_A+R}=0.  
\ee
In principle one could search for an exact solution $R(\Phi_A)$ which  seems however quite challenging. Therefore, as  in standard field theory, we can resort to perturbation theory.
We setup a perturbative approach by introducing a coupling constant $g$ and we write
\be  
\Phi=\Phi(g)=g \Phi_A+\sum_{n=2}^\infty g^n\ R_n.
\ee
The $R$ equations can now be solved iteratively by expanding in powers of $g$. The first non-trivial equations are given by
\be
\eta_0 Q_B R_2&=&\bar P_0\ \frac12[\eta_0\Phi_A,Q_B\Phi_A]\\
\eta_0 Q_BR_{3}&=&\bar P_0{\Bigg(}\dfrac{1}{2} [\eta_0 \Phi_A , Q_BR_{2}] +\dfrac{1}{2} [\eta_0 R_{2},Q_B\Phi_A] \\
&&\quad\ - \dfrac{1}{3!} [\eta_0 \Phi_A,[\Phi_A,Q_B\Phi_A]] + \dfrac{1}{12}[\Phi_A,[\eta_0 \Phi_A , Q_B\Phi_A]] {\Bigg)}\0\\
\eta_0 Q_B R_4&=&\bar P_0 {\Big(}\cdots{\Big)},
\ee
where, since in this paper we will only deal with these equations at order $g^2$, we do not write down the equation at order 4 and higher.
These equations can all be solved by fixing the gauge $\xi_0=b_0=0$. In particular at order $g^2$ we find
\be\label{sol2}
R_2=-\frac12 \xi_0\ \frac{b_0}{L_0}\bar P_0 [\eta_0\Phi_A,Q_B\Phi_A].
\ee
Notice that the presence of the projector outside of the kernel of $L_0$  ensures that the action of $b_0/L_0$ is well-defined.

By plugging \eqref{sol2} into the classical action \eqref{action} we get the tree level effective action at order $g^4$ which explicitly reads\footnote{We are not interested in $O(g^3)$ terms which, in the present setting, come exclusively from the elementary cubic vertex in \eqref{action} and which are not affected by integrating out the heavy fields. Moreover in the cases we analize these couplings are identically vanishing due to zero momentum }
\be
S_{\rm eff}[\Phi_A]&=& -g^4\dfrac{1}{2}\Tr[(Q_B \eta_0 R_{2}) \, R_{2}]-g^4\dfrac{1}{2} \Tr[R_{2}\,[\eta_0 \Phi_A,Q_B\Phi_A]] \0\\
& & -g^4 \dfrac{1}{4!}\Tr[(\eta_0 \Phi_A) \,[\Phi_A ,\,[\Phi_A, Q_B \Phi_A]]] + O(g^5)\0\\
&=& g^4\dfrac{1}{8}\Tr\left[[\eta_0 \Phi_A,Q_B\Phi_A]\, \xi_0 \dfrac{b_0}{L_0}\bar{P} \, [\eta_0 \Phi_A,Q_B\Phi_A]\right]\0 \\
& & -g^4\dfrac{1}{24}\Tr[[\eta_0 \Phi_A, \Phi_A] \,[\Phi_A, Q_B \Phi_A]]+O(g^5).\label{seff4}
\ee 
Notice that the effective action to this order contains the elementary quartic vertex of \eqref{action} plus a term with a propagator, which is the result of having integrated out the heavy fields. In case of the zero momentum gauge field \eqref{gaugefield} this quantity has been computed exactly by Berkovits and Schnabl and shown to reproduce the expected quartic Yang-Mills potential \cite{BS}
\be
\Phi_A&=&A_{\mu} c\gamma^{-1}\psi^{\mu}\0\\
&\downarrow&\0\\
S_{\rm eff}[\Phi_A]&=&g^4\left( - \dfrac{1}{4} \, \mathrm{tr} \left[A^2 A^2 \right] - \dfrac{1}{8} \, \mathrm{tr} \left[A_\mu A_\nu A^\mu A^\nu \right] \right)_{\rm propagator}\0\\&&+g^4\left(- \dfrac{1}{8} \, \mathrm{tr} \left[A_\mu A_\nu A^\mu A^\nu \right] + \dfrac{1}{2} \, \mathrm{tr} \left[A^2 A^2 \right]\right)_{\rm 4-vertex} \0 \\
&=& - \dfrac{g^4}{8} \, \mathrm{tr} \left[[A_\mu ,A_\nu] [A^\mu, A^\nu ]\right] ,\,\label{YM}
\ee
where we have distinguished the contributions from the propagator term and the elementary quartic vertex in \eqref{seff4}. 

\subsection{Subtleties from picture changing}
Although it is not apparent in the explicit expression \eqref{seff4}, this quantity is affected by some subtleties. 
These subtleties are easiest to  see by taking the explicit example of the zero momentum gauge field \eqref{gaugefield} but, as we will see later on, they are in fact fairly generic.
Let's  have a closer look at the propagator term in \eqref{seff4}
\be
S^{(4)}_{\rm prop}&=&\dfrac{1}{8}\Tr\left[[\eta_0 \Phi_A,Q_B\Phi_A]\, \xi_0 \dfrac{b_0}{L_0}\bar{P}_0 \, [\eta_0 \Phi_A,Q_B\Phi_A]\right]\label{Sprop}
\ee
 Given this expression we can ``change'' the picture assignment  integrating by parts the derivations $Q_B$ and $\eta_0$, in a way analogous to \cite{ONT}. Taking advantage of $\eta_0 Q_B\Phi_A=0$, this is done as follows
 \be
 S^{(4)}_{\rm prop}&=&\dfrac{1}{8}\Tr\left[[\eta_0 \Phi_A,Q_B\Phi_A]\, \xi_0 \dfrac{b_0}{L_0}\bar P_0\, [\eta_0 \Phi_A,Q_B\Phi_A]\right]\0\\
 &=&\dfrac{1}{8}\Tr\left[\left(\eta_0[ \Phi_A,Q_B\Phi_A]\right)\, \xi_0 \dfrac{b_0}{L_0}\bar P_0 \, \left(Q_B[ \Phi_A,\eta_0\Phi_A]\right)\right]\0\\
 &=&\dfrac{1}{8}\Tr\left[[ \Phi_A,Q_B\Phi_A]\,  \dfrac{b_0}{L_0}\bar P_0 \, \left(Q_B[ \Phi_A,\eta_0\Phi_A]\right)\right]\0\\
 &=&-\dfrac{1}{8}\Tr\left[[ Q_B\Phi_A,Q_B\Phi_A]\,  \dfrac{b_0}{L_0}\bar P_0 \, [ \Phi_A,\eta_0\Phi_A]\right]\0\\
 &&+\dfrac{1}{8}\Tr\left[\left([ \Phi_A,Q_B\Phi_A]\right)\, \left[ \dfrac{b_0}{L_0}\bar P_0, Q_B\right]\,[ \Phi_A,\eta_0\Phi_A]\right]\0\\
 &=&\dfrac{1}{8}\Tr\left[[ Q_B\Phi_A,Q_B\Phi_A]\, \xi_0 \dfrac{b_0}{L_0}\bar P_0 \, [\eta_0 \Phi_A,\eta_0\Phi_A]\right]\0\\
 &&+\dfrac{1}{8}\Tr\left[[ \Phi_A,Q_B\Phi_A]\, (1-P_0) [ \Phi_A,\eta_0\Phi_A]\right].\label{prop-pc}
  \ee
Notice that we have taken into account the nontrivial projector outside the kernel of $L_0$ that arises in the commutator
\be
\left[Q_B,\frac{b_0}{L_0}\bar P_0\right]=\bar P_0=1-P_0.
\ee
Therefore we see that the performed manipulation (which is obviously related to picture-changing) gives rise to the singular term\footnote{This term was ignored in \cite{ONT}. This is justified at generic momentum, but not at zero momentum, where the propagator has to be defined avoiding the kernel of $L_0$.}
\be
-\dfrac{1}{8}\Tr\left[[ \Phi_A,Q_B\Phi_A]\, P_0 [ \Phi_A,\eta_0\Phi_A]\right] = + \dfrac{1}{8}\Tr\left[[  \Phi_A,\eta_0 \Phi_A] \, P_0 [ \Phi_A,Q_B\Phi_A]\right],\label{proj}
\ee
which we now analyze. We start by considering
\be
P_0[\Phi_A,Q_B\Phi_A]&=&P_0 U_3^* \left(\Phi_A \left(\frac1{\sqrt3}\right)Q_B\Phi_A\left(-\frac1{\sqrt3}\right)-Q_B\Phi_A \left(\frac1{\sqrt3}\right)\Phi_A\left(-\frac1{\sqrt3}\right)\right)\ket0.\0
\ee
In the above equation we have represented the star product as the action of the $U$ operators on the finite distance OPE, as discussed in \cite{wedges}. The action of $P_0$ only selects the level zero contribution from the above wedge with insertions. This means that only the level zero component of the $U$ operator (that is the identity) gives contribution and, moreover, only the level zero contribution from the OPE (if present) participates. It is not difficult to check that for the explicit case of the gauge field \eqref{gaugefield} we 
 get
\be
P_0[\Phi_A,Q_B\Phi_A] &=& + A_\mu A_\nu \, c \gamma^{-1} \psi^\mu \left(\frac1{\sqrt3}\right)  \left[ c \left(i \sqrt{2}\partial X^\nu \right) - \gamma \psi^\nu \right] \left(- \frac1{\sqrt3}\right) \0\\
&& - A_\mu A_\nu \,\left[ c \left(i \sqrt{2}\partial X^\nu \right) - \gamma \psi^\nu \right] \left(\frac1{\sqrt3}\right) \, c \gamma^{-1} \psi^\mu \left(- \frac1{\sqrt3}\right) {\Bigg |}_{0}\0\\
&=& -[A_\mu,A_\nu]c\,\psi^{\mu}\psi^{\nu}(0)\ket0.\label{proj-ope} 
\ee
Notice that the matter primaries $\psi$ and $\partial X$ have a regular OPE and so their product doesn't contribute at level zero. Moreover the projector selects the terms where super-ghosts OPE closes on the identity in the form 
\be
\gamma(z)\gamma^{-1}(w)=1+O(z-w),
\ee
and because of the special role played by  $\gamma^{-1}$, this is a large Hilbert space phenomenon.
In the same fashion we can check that
\be
P_0[\Phi_A,\eta_0 \Phi_A]=[A_\mu,A_\nu]\xi\,  c\del c\, e^{-2\phi}\, \psi^{\mu} \psi^{\nu}(0)\ket0.
\ee
From these simple considerations we see that the singular projector term \eqref{proj}, far from vanishing, is in fact proportional to the full Yang-Mills quartic potential
\be
\Tr\left[[  \Phi_A,\eta_0 \Phi_A] \, P_0 [ \Phi_A,Q_B\Phi_A]\right] \sim \tr\left[[A_\mu,A_\nu][A^\mu,A^\nu]\right].\label{proj-comm}
\ee
Notice that this quantity involves the computation of a four-point function in a region of moduli space where two insertions
are collapsed on one-another and replaced by their level zero contribition in the OPE. Equivalently we can think of $P_0$ as an infinitely long Siegel-gauge strip
\be
P_0=\lim_{t\to\infty}e^{-t L_0},
\ee
 which creates a degenerated four-punctured disk at the boundary of moduli space.
Taking into account this boundary term in the ``picture-changed'' propagator term \eqref{prop-pc}, one can repeat (with the obvious modifications) Berkovits-Schnabl computation \cite{BS} and get the correct Yang-Mills quartic potential \eqref{YM}.

\section{Localization of the action at quartic order}\label{sec3}

Given the fact that \eqref{proj-comm} is already proportional to the full answer, it is tempting to claim that   the Yang-Mills quartic potential is fully localized at the boundary of moduli space. However 
it doesn't seem to be possible to prove this with the ingredients we have used up to now. Recently Sen has provided a similar mechanism for the heterotic string, in the small Hilbert space, \cite{Sen-restoration} which is based on the existence of a conserved charge in the matter sector. In the sequel we will also assume an extra conserved charge in the matter sector and things will drastically simplify.
\subsection{Conserved charge and $\mathcal{N}=2$}

 This charge, which we will call $J$, is understood as a  $U(1)$ $R$-symmetry of an
$\mathcal{N}=2$  world-sheet supersymmetry in the matter sector. It is well known that the original $\mathcal{N}=1$ worldsheet supersymmetry is  enhanced to an $\mathcal{N}=2$ in  superstring backgrounds supporting space-time fermions (and thus space-time supersymmetry), see for example \cite{susy2}. In order for the $\mathcal{N}=1\to \mathcal{N}=2$ enhancement to happen, it must be possible to express the original $\mathcal{N}=1$ matter supercurrent $T_F$ as the sum of two supercurrents of opposite ``chirality"
\be
T_F=T_F^{(+)}+T_F^{(-)},\label{TFpm}
\ee
so that an $\mathcal{N}=2$ super-Virasoro algebra can be realized
\be
T(z)\ T(w)&=&\frac{c/2}{(z-w)^4}+\frac{2T(w)}{(z-w)^2}+\frac{\del T(w)}{z-w}+...\\
T(z)\ T_F^{(\pm)}(w)&=&\frac32\frac{ T_F^{\pm}(w)}{(z-w)^2}+\frac{\del T_F^{(\pm)}(w)}{z-w}+...\\
T_F^{(+)}(z)\ T_F^{(-)}(w)&=&\frac{2c/3}{(z-w)^3}+\frac{J(w)}{(z-w)^2}+\frac1{z-w}(2T(w)+\del J(w))+...\\
T(z)\ J(w)&=&\frac{J(w)}{(z-w)^2}+\frac{\del J(w)}{z-w}+...\\
J(z)\ T_F^{(\pm)}(w)&=&\pm\frac{ T_F^{(\pm)}(w)}{z-w}+...\\
J(z)\ J(w)&=&\frac{c/3}{(z-w)^2}+...
\ee
The full matter SCFT has $c=15$ but one may be interested in subsector of the full background (a flat 4-dimensional Minkowski space $(c=6)$ or a Calabi-Yau internal space $(c=9)$ are both examples  with enhanced $\mathcal{N}=2$).
We will be interested in the case where our original $\mathcal{N}=1$ superconformal primary $\VV_{\frac12}$ \eqref{stringfield} splits into the sum of two  ``short" $\mathcal{N}=2$ superconformal primaries 
\be
\VV_{\frac12}=\VV_{\frac12}^{(+)}+\VV_{\frac12}^{(-)},
\ee

such that
\be
T_F^{(\pm)}(z)\VV^{(\mp)}_{\frac12}(w)&=&\frac1{z-w}\VV^{(\mp)}_{1}(w)+...\label{V1pm}\label{descpm}\\
T_F^{(\mp)}(z)\VV^{(\mp)}_{\frac12}(w)&=&\textrm{regular}.
\ee
The $R$-current $J(z)$  defines a conserved charge
\be
J_0=\oint\frac{dz}{2\pi i}\,J(z),
\ee
and the short superconformal primaries $\VV_{\frac12}^{(\pm)}$ are $J_0$-eigenstates 
\be
J_0 \VV_{\frac12}^{(\pm)}=\pm  \VV_{\frac12}^{(\pm)}.
\ee
From (\ref{TFpm},\ref{V1pm}) we see that the original matter field $\VV_1$ also decomposes as
\be
\VV_1=\VV_{1}^{(+)}+\VV_{1}^{(-)}.
\ee
However, despite the notation, the super-descendents $\VV_1^{\pm}$ are not charged under $J_0$, because the net $J$-charge in \eqref{descpm} is zero.
In the matter SCFT only correlators with total vanishing $J$-charge are non-zero. This gives a selection rule that drastically simplifies the computation of the effective action \eqref{seff4}
\subsection{Localization of the effective action}
As anticipated above, we now assume that our physical string field $\Phi_A$ decomposes in eigenstates with  $J$ charge equal to $\pm1$
\be
\Phi_A&=&\Phi_A^{(+)}+\Phi_A^{(-)},
\ee
with
\be
\Phi_A^{(\pm)}=c\gamma^{-1}\VV_{\frac12}^{(\pm)}.
\ee
Now we want to compute the effective action \eqref{seff4} in the presence of the above decomposition
\be
S_{\rm eff}^{(4)}(\Phi_A)=S_{\rm eff}^{(4)}\left(\Phi_A^{(+)}+\Phi_A^{(-)}\right).
\ee
The details of the computations are shown in the appendix \ref{appA}. We just report here the final result which gives us a completely localized effective action where only projector-type terms remain
\be
S_{\rm eff}^{(4)}(\Phi_A)&=&\frac18 \Trr{[\Phi_A^{(-)},\eta_0\Phi_A^{(-)}]\,P_0\,  [\Phi_A^{(+)},Q_B\Phi_A^{(+)}] + [\Phi_A^{(+)},\Phi_A^{(-)}]\, P_0 [\eta_0\Phi_A^{(-)},Q_B\Phi_A^{(+)}]}\0\\
&& + \left(\Phi_A^{(+)}\leftrightarrow \Phi_A^{(-)}\right)\label{dff}\\
&=&\frac18\left[\left\langle\widehat h^{(--)}{\Big |}h^{(++)}\right\rangle+\left\langle\widehat g^{(+-)}{\Big |}g^{(-+)}\right\rangle+{\Big (}+\leftrightarrow-{\Big)}\right],
\ee
which shows that the quartic effective action is entirely given by two-point functions of Fock space states which we now analyse.

\subsection{Auxiliary fields}

The basic fields which enter in the above expression for the effective action are\footnote{It is important to note that while $P_0[\eta_0\Phi_A,Q_B\Phi_A]$ is identically vanishing, when we consider \eqref{auxx3} and \eqref{auxx4} we find instead a non-vanishing contribution due to the different Chan-Paton structure. In particular we have that $g^{(\pm\mp)}=-g^{(\mp\pm)}$ and $\widehat g^{(\pm\mp)}=-\widehat g^{(\mp\pm)}$.}
\begin{eqnarray}
h^{(\pm\pm)}&\equiv& P_0[\Phi_A^{(\pm)}, Q_B\Phi_A^{(\pm)}]\quad \textrm{with}\ J=\pm2\\
\widehat h^{(\pm\pm)}&\equiv& P_0[\Phi_A^{(\pm)}, \eta_0\Phi_A^{(\pm)}]\quad \textrm{with}\ J=\pm2\\
g^{(\pm\mp)}&\equiv& P_0[\eta_0\Phi_A^{(\pm)},Q_B\Phi_A^{(\mp)}]\quad  \textrm{with}\ J=0 \label{auxx3}  
\\
\widehat g^{(\pm\mp)}&\equiv& P_0[\Phi_A^{(\pm)},\Phi_A^{(\mp)}]\quad \textrm{with}\ J=0\label{auxx4}.
\end{eqnarray}
To determine the explicit form of the above string fields we only need to know the leading OPE between the matter superconformal primaries
\be
\VV_{\frac12}^{(\pm)}(z)\VV_{\frac12}^{(\pm)}(-z)&=&\HH_1^{(\pm)}(0)+\cdots \label{aux1}\\
\VV_{\frac12}^{(\mp)}(z)\VV_{\frac12}^{(\pm)}(-z)-\VV_{\frac12}^{(\pm)}(z)\VV_{\frac12}^{(\mp)}(-z)&=&\frac1{2z}\HH_0+\cdots. \label{aux2}
\ee
Where we have introduced the ``auxiliary fields" $\HH_1^{(\pm)}$ which are weight-one matter primaries with $J$ charge equal to $\pm2$ and $\HH_0$ which is proportional to the identity in the matter CFT and is neutral under $J$. 
Then, using the universal OPE's in the ghost/superghost sector 
\be
c(z) \, c(-z) \sim - 2z \,c \partial c(0) +\cdots \qquad, \qquad  \xi(z) \, \xi(-z) \sim - 2z \, \xi \partial \xi(0) +\cdots
\ee
\be
e^{-\phi} (z) \, e^{-\phi} (-z) \sim \dfrac{1}{2z}e^{-2 \phi}(0) +\cdots \qquad, \qquad \gamma(z) \gamma^{-1}(-z) \sim 1+\cdots
\ee
we find by a direct computation analogous to \eqref{proj-ope}
\be
h^{(\pm\pm)}&=&-2c\,\HH_{1}^{(\pm)}\\
\widehat h^{(\pm\pm)}&=& 2c\del c\, \xi \, e^{-2\phi}\, \,\HH_{1}^{(\pm)}\\
g^{(\pm\mp)}&=& \mp c\, \eta \, \HH_0 \label{aux3}\\   
\widehat g^{(\pm\mp)}&=&  \mp c\del c\, \xi\del\xi\, e^{-2\phi}\, \HH_0. \label{aux4}
\ee
Now all the contribution to the effective action just depends on the auxiliary fields  \eqref{aux1}, \eqref{aux2} and their two-point functions.

Notice in particular that in the case that the OPE structure of the matter fields $\VV_{\frac12}^{(\pm)}$ is not precisely of the form (\ref{aux1},\ref{aux2}) (which is the case for example at generic momenta), the amplitude \eqref{dff} would identically vanish.

\section{Examples}\label{sec4}

To simplify a bit our result we can compute the universal contributions from the ghosts  as
\be
\langle c\del c\, \xi \, e^{-2\phi}(z) \,c(w) \rangle &=& -(z-w)^2 , \\
\langle c\del c\, \xi \del\xi \, e^{-2\phi}(z) \,c \, \eta (w)\rangle &=& -1  , 
\ee
remaining with purely matter two-point functions
\be
S_{\rm eff}^{(4)}(\Phi_A)=\tr\left[\bra{\HH_{1}^{(+)}}\,\HH_1^{(-)}\rangle + \frac14\bra{\HH_{0}}\,\HH_0\rangle \right],\label{seffmatt}
\ee
where $\tr$ is the trace in Chan-Paton space and the bracket is in the matter sector.
This compact result is universal. The details of the $\HH$ fields depend on the chosen matter SCFT, i.e. the string background in which we are interested. To illustrate how this mechanism works we now give two concrete examples.
\subsection{Yang-Mills}

Consider a system of $N$ coincident $D(2n)$ euclidean branes. Their worldsheet theory contains the $\psi^{\mu}$ superconformal fields which we rearrange according to a $U(n)\in SO(2n)$ decomposition
\be
\psi^{\j}&=&\frac1{\sqrt2}(\psi^{2j-1}+i\psi^{2j})\0\\
\psi^{\bj}&=&\frac1{\sqrt2}(\psi^{2j-1}-i\psi^{2j}),\label{psicomm}
\ee
 with $(\j,\bj,j)=1,...,n$.
We can bosonize these fields with $n$ free bosons $h_i$ such that
\be
\psi^{\j}&=&e^{i h_j},\\
\psi^{\bj}&=&e^{-i h_j}.
\ee
The localizing $R$-charge can be taken to be\footnote{We have at our disposal $n$ decoupled $N=2$ SCFT, so we can choose any linear combinations of the individual $R$-charges $\del h_i$. We choose this particular combination for definitness.}
\be
J(z)=-i\sum_{j=1}^{n} \del h_j(z). \label{current}
\ee
With this choice we have
\be
J_0\psi^{\j}&=&+\psi^{\j},\\
J_0\psi^{\bj}&=&-\psi^{\bj}.
\ee
We then write
\be
\Phi_A=A_\mu c\gamma^{-1}\psi^{\mu}=\Phi_A^{(+)}+\Phi_A^{(-)},
\ee
with\footnote{We are in euclidean space and we don't distinguish between upper and lower indices.}
\be
\Phi_A^{(+)}&=&A_{\j}\,  c\gamma^{-1}\psi^{\j},\\
\Phi_A^{(-)}&=&A_{\bj}\,  c\gamma^{-1}\psi^{\bj}.
\ee
Therefore our matter superconformal primaries of definite $J$-charge are given by
\be
\VV^{(+)}_{\frac12}(z)&=&A_{\j}\,\psi^{\j}(z),\\
\VV^{(-)}_{\frac12}(z)&=&A_{\bj}\,\psi^{\bj}(z),
\ee
and the auxiliary fields are easily extracted from the leading term in the OPE
\be
\HH_1^{(+)}(z)&=&\lim_{\epsilon\to0}\VV^{(+)}(z+\epsilon)\VV^{(+)}(z-\epsilon)=\frac12[A_{\i},A_{\j}]\, \psi^{\i \j}(z),\\
\HH_1^{(-)}(z)&=&\lim_{\epsilon\to0}\VV^{(-)}(z+\epsilon)\VV^{(-)}(z-\epsilon)=\frac12[A_{\bi},A_{\bj}]\, \psi^{\bi \bj}(z),\\
\HH_0(z)&=&\lim_{\epsilon\to0}(2\epsilon)\VV^{(-)}(z+\epsilon)\VV^{(+)}(z-\epsilon)=[A_{\bj},A_{\j}].
\ee
where $\psi^{\i \j}(z) \equiv : \psi^{\i} \psi^{\j}: (z)$. The effective action \eqref{seffmatt} is then easily computed to be
\be
S_{\rm eff}^{(4)}(\Phi_A)&=&\tr\left[\bra{\HH_{1}^{(+)}}\,\HH_1^{(-)}\rangle+\frac14\bra{\HH_{0}}\,\HH_0\rangle \right]\0\\
&=&\tr\left[-\frac12[A_{\i},A_{\j}][A_{\bi},A_{\bj}]+\frac14[A_{\bj},A_{\j}][A_{\bi},A_{\i}]\right].
\ee
To recover a more familiar covariant expression, notice that the  $N\times N$ matrices $A_{\j}, A_{\bj}$ are related to the original $A_\mu$'s by \eqref{psicomm}
\be
A_{\j}&=&\tau_{\j}^\mu\, A_\mu\\
A_{\bj}&=&\bar \tau_{\bj}^\mu\, A_\mu,
\ee
where the only non-vanishing entries of $\tau$ and $\bar\tau$ are 
\be
\tau^{2j-1}_{\j}&=&\frac1{\sqrt2}=\bar\tau^{2j-1}_{\bj}\\
\tau^{2j}_{\j}&=&\frac i{\sqrt2}=-\bar\tau^{2j}_{\bj}.
\ee
We can easily check that
\be
\sum_{\j=1}^n\, \tau^{\mu}_{\j}\bar\tau^{\nu}_{\bj}=\frac12\left(\delta^{\mu\nu}-i\epsilon^{\mu\nu}\right),
\ee
where $\epsilon^{\mu\nu}=\epsilon_{\mu\nu}$ is a block-diagonal anti-symmetric matrix whose only non-zero entries are given by
\be
\epsilon_{2j-1,\, 2j}=-\epsilon_{2j,\, 2j-1}=1,\quad\quad j=1,\cdots,n.
\ee
Using these properties, together with the ciclicity of the trace, one can easily verify that the usual covariant form of the Yang-Mills potential is reproduced
\be
S_{\rm eff}^{(4)}(\Phi_A)&=&\tr\left[-\frac12[A_{\i},A_{\j}][A_{\bi},A_{\bj}]+\frac14[A_{\bj},A_{\j}][A_{\bi},A_{\i}]\right]\0\\
&=& - \frac18 \tr\left[[A_\mu,A_\nu][A^\mu,A^\nu]\right].
\ee
The quartic potential of the scalars transverse to the $D(2n)$ branes and their interaction with the gauge fields can be obtained in the same way.\footnote{These actions are simple dimensional reductions of 10D SYM}

\subsection{ $D3$/$D(-1)$ system}

We now consider the low energy effective action of a system of $N$ coincident (euclidean) D3 branes with $k$ $D(-1)$ branes sitting on the $D3$ world-volume. This system is known to give a string theory description of supersymmetric gauge theory instantons  \cite{Witten:1995im,  Douglas:1996uz}. A direct string theory construction of the $D3$--$D(-1)$ effective action to leading order in $\alpha'$ has been given in\cite{torinesi}. Here we will be interested to show that the effective action of this system is also exactly localized at the boundary of moduli space.  \\
The presence of the $D3$ branes brakes  $SO(10)$ to the product of a Wick rotated Lorentz group $SO(4)$ on the $D3$ branes and  $SO(6)$ along the transverse directions. Then we have three sectors of open strings:
\begin{itemize}
\item $D3$--$D3$ strings. These are strings with both endpoints on the set of $D3$ branes. Their description in terms of string fields is exactly the same we have used in the previous sections. These string fields carry an $N\times N$ Chan-Paton factor.
\item $D(-1)$--$D(-1)$ strings. These are strings with both endpoints on the set of $D(-1)$ branes. The  corresponding string fields are exactly the same, up to the Chan-Paton factor that now is a $k \times k$ matrix.
\item $D3$--$D(-1)$ strings. These are stretched strings between the two types of $D$-branes. The corresponding vertex operators must include  twist fields ($\Delta$,  $\bar\Delta$) that  are necessary to change the boundary conditions from Neumann to Dirichlet  ($\Delta$) and viceversa ($\bar\Delta$) in the $X^\mu$ sector. These are a pair of conjugated boundary primary fields of weight $\frac14$.  The bosonic twist fields $\Delta, \bar\Delta$ must be dressed with four-dimensional spin fields $S^\alpha$ (or $S^{\dot\alpha}$ in case of anti $D(-1)$'s) in order to change the boundary conditions of the $\psi^\mu$ system. See \cite{torinesi} for further details  (which are partly summarized in appendix \ref{appB}).The Chan-Paton factor carried by these string fields is a $N \times k$ or a $k \times N$ matrix.

It is important to notice that the composite fields $\Delta S^\alpha$ and $\bar\Delta S^\alpha$  have analogous properties to the worldsheet fermion $\psi$, in particular they are superconformal primaries of weight $\frac12$.  \end{itemize}
Assembling things together, the total  massless string field is then given by
\begin{equation}
\Phi_A (z)=c \gamma^{-1} \VV_\frac{1}{2}(z)= c \gamma^{-1} \left(\begin{matrix}
A & \omega \\
\bar{\omega} & a
\end{matrix}\right) (z),
\end{equation}
where
\begin{equation}
A(z)=A_{\mu} \psi^{\mu}(z) + \phi_p  \psi^{p}(z),
\end{equation}
\begin{equation}
\omega(z) = \omega^{N \times k}_{\alpha}  \, \Delta S^{\alpha}(z),
\end{equation}
\begin{equation}
\bar{\omega}(z) = \bar{\omega}^{k \times N}_{\alpha} \, \bar{\Delta} S^{\alpha}(z),
\end{equation}
\begin{equation}
a(z)=a_{\mu}  \psi^{\mu}(z) + \chi_p  \psi^{p}(z).
\end{equation}
Greek indices $\mu$ label the directions along the D3 branes, roman indices $p$ label the transverse directions while four-dimensional spinor indices $\alpha$ are $(\frac12,\frac12)$ and $(-\frac12,-\frac12)$. 

This string field can be decomposed in a basis of eigenstates of the current $J_0$ as defined in \eqref{current}
\be
J_0=-i\sum_{k=1}^{5} \oint \frac{dz}{2\pi i} \del h_k(z).\label{J0}
\ee

The diagonal matter vertex is easily decomposed as 
\be
A = A^{(+)}+A^{(-)} + \phi^{(+)}+\phi^{(-)} \qquad , \qquad a = a^{(+)}+a^{(-)} + \chi^{(+)}+\chi^{(-)}
\ee
where using the same notation as in \eqref{psicomm}
\be
A^{(+)} =  A_{\j}\, \psi^{\j} \qquad , \qquad  A^{(-)} =  A_{\bar{\j}}\, \psi^{\bar{\j}}, 
\ee
\be
\phi^{(+)} =  \phi_{m}\, \psi^{m} \qquad , \qquad  \phi^{(-)} =  \phi_{\bar{m}}\, \psi^{\bar{m}}, 
\ee
\be
a^{(+)} =  a_{\j}\, \psi^{\j} \qquad , \qquad  a^{(-)} =  a_{\bar{\j}}\, \psi^{\bar{\j}}, 
\ee
\be
\chi^{(+)} =  \chi_{m}\, \psi^{m} \qquad , \qquad  \chi^{(-)} =  \chi_{\bar{m}}\, \psi^{\bar{m}}.
\ee
Here $\j=1,2$ and $\bar{\j} = \bar{1},\bar{2}$ indices denote respectively the fundamental  and antifundamental representation of $SU(2)\subset SO(4)$, while $m=1,2,3$ and $\bar{m}= \bar{1}, \bar{2}, \bar{3}$ indices denote respectively the  fundamental and antifundamental representation  of $SU(3)\subset SO(6)$. 

The off-diagonal spin fields are defined through bosonization by the scalars $h_1, h_2$ and we consider
\be
S^{(\frac12,\frac12)} = e^{\frac{i}{2} \left( h_1 + h_2 \right)} \qquad, \qquad S^{(-\frac12,-\frac12)} = e^{- \frac{i}{2} \left( h_1 + h_2 \right)}
\ee
such that \footnote{Other choices for the localizing current are in principle possible as in the case of pure Yang-Mills. For example another possible choice of $J$ along the $D3$ branes is the linear combination of $\partial h_i$ with a relative minus sign between $\del h_1$ and $\del h_2$. However, while for the pure Yang-Mills all the possible choices are equivalent, the presence of the stretched strings makes a difference whether we have anti $D(-1)$'s rather than $D(-1)$'s. In presence of anti $D(-1)$'s the corresponding spin fields would have opposite chirality and they would  be un-charged under $J_0$ defined in \eqref{J0}. So to localize their action one would choose a different $J_0$ with an opposite sign in front of $\del h_2$. } 

\be
J_0 \, S^{(\frac12,\frac12)} = + S^{(\frac12,\frac12)} \qquad , \qquad J_0 \, S^{(-\frac12,-\frac12)} = - S^{(-\frac12,-\frac12)}.
\ee

For more details on bosonization and spin fields, see  appendix \ref{appB}. Finally we can decompose the off-diagonal part of $\VV_{\frac12}$ as
\be
\omega = \omega^{\left( +\right) } + \omega^{\left( -\right) } \qquad , \qquad  \bar{\omega} = \bar{\omega}^{\left( +\right) } + \bar{\omega}^{\left( -\right) }
\ee
where 
\be
\omega^{\left( +\right) } = \omega_1 \, \Delta S^{(\frac12,\frac12)} \qquad , \qquad \omega^{\left( -\right) } = \omega_2 \, \Delta S^{(-\frac12,-\frac12)}
\ee
\be
\bar{\omega}^{\left( +\right) } = \bar{\omega}_1 \, \bar\Delta S^{(\frac12,\frac12)} \qquad , \qquad \bar{\omega}^{\left( -\right) } = \bar{\omega}_2 \, \bar\Delta S^{(-\frac12,-\frac12)}.
\ee
Summarising
\be
\Phi_A (z) = \Phi_A^{(+)}(z)+\Phi_A^{(-)}(z),
\ee
where
\be
\Phi_A^{(+)} (z) = c \gamma^{-1}                                 \left(\begin{matrix}
A^{\left( +\right) } + \phi^{\left( +\right) } & \omega^{\left( +\right) } \\
\bar{\omega}^{\left( +\right) } & a^{\left( +\right) } + \chi^{\left( +\right) }
\end{matrix}\right)(z),
\ee
\be
 \Phi_A^{(-)} (z) =c \gamma^{-1}                                 \left(\begin{matrix}
A^{\left( -\right) } + \phi^{\left( -\right) } & \omega^{\left( -\right) } \\
\bar{\omega}^{\left( -\right) } & a^{\left( -\right) } + \chi^{\left( -\right) }
\end{matrix}\right)(z).
\ee
Then using our general result \eqref{seffmatt} the quartic potential is given by the two-point functions of the matter auxiliary fields \eqref{aux1}, \eqref{aux2}
 $\HH_{0} , \HH_{1}^{\left( +\right)}  ,  \HH_{1}^{\left( -\right) }$. The computations are easily carried out using standard OPE's and for the sake of clarity we write the auxiliary fields as a sum of two components, the first one along the D3 branes directions
\be
\HH_{1}^{\left( +\right) D3} = \left(\begin{matrix}
 \left[  A_1 ,A_2 \right]  +  \omega_1 \bar{\omega}_1 & 0 \\
0 & \left[ a_1 ,a_2 \right] - \bar{\omega}_1 \omega_1
\end{matrix}\right) \psi_{1\,2} \vert 0 \rangle, \,  \label{HD3+}
\ee
\be
\HH_{1}^{\left( -\right) D3} = \left(\begin{matrix}
 \left[  A_{\bar{1}} ,A_{\bar{2}} \right]  +  \omega_2 \bar{\omega}_2 & 0 \\
0 & \left[ a_{\bar{1}} ,a_{\bar{2}} \right] - \bar{\omega}_2 \omega_2
\end{matrix}\right) \psi_{\bar{1}\, \bar{2}}  \vert 0 \rangle, \, \label{HD3-}
\ee
\be
\HH^{D3}_{0} = \left(\begin{matrix}
 \left[  A_{\bar{\j}} ,A_{\j} \right]  - \left(   \omega_1 \bar{\omega}_2 + \omega_2 \bar{\omega}_1 \right) & 0 \\
0 & \left[  a_{\bar{\j}} ,a_{\j} \right]  + \left(  \bar \omega_1 \omega_2 +\bar \omega_2\omega_1 \right) 
\end{matrix} \right) \vert 0 \rangle, \, \label{HD30}
\ee
and the second one along the transverse directions
\be
 \HH_{1}^{\left( +\right) T} = \left(\begin{matrix}
 \dfrac{1}{2}\left[  \phi_m ,\phi_n \right] \ \psi^{m\, n}  + \left[  A_{\j} ,\phi_m \right] \ \psi^{\j\, m}  & \left( \phi_m \omega_1 - \bar{\omega}_1 \chi_m \right)\   \psi^m \Delta S^{(\frac12,\frac12)}  \\
\left( \chi_m\bar{\omega}_1 - \bar{\omega}_1 \phi_m\right) \  \psi^m \bar{\Delta} S^{(\frac12,\frac12)}  & \dfrac{1}{2}\left[  \chi_m ,\chi_n \right]\  \psi^{m\, n}  + \left[  a_{\j} ,\chi_m \right]\  \psi^{\j\, m}
\end{matrix}\right) \vert 0\rangle, \, \0  \\ \label{H+}
\ee
\be
 \HH_{1}^{\left( -\right)T} = \left(\begin{matrix}
 \dfrac{1}{2}\left[  \phi_{\bar{m}} ,\phi_{\bar{n}} \right]\  \psi^{\bar{m}\, \bar{n}}  + \left[  A_{\bar{\j}} ,\phi_{\bar{m}} \right] \ \psi^{\bar{\j}\, \bar{m}}  & \left( \phi_{\bar{m}} \omega_2 - \bar{\omega}_2 \chi_{\bar{m}} \right)\  \psi^{\bar{i}} \Delta S^{(-\frac12,-\frac12)}  \\
\left( \chi_{\bar{m}} \bar{\omega}_2 - \bar{\omega}_2 \phi_{\bar{m}}\right) \  \psi^{\bar{m}} \bar{\Delta} S^{(-\frac12,-\frac12)}  & \dfrac{1}{2} \left[  \chi_{\bar{m}} ,\chi_{\bar{n}} \right] \psi^{\bar{m} \, \bar{n}}  + \left[  a_{\bar{\j}} ,\chi_{\bar{m}} \right] \psi^{\bar{\j}\, \bar{m}}
\end{matrix}\right) \vert 0\rangle, \,  \0 \\ \label{H-}
\ee
\be
\HH^T_{0} = \left(\begin{matrix}
 \left[  \phi_{\bm} ,\phi_{m} \right]  & 0 \\
0 & \left[ \chi_{\bar{m}} ,\chi_{m} \right] 
\end{matrix}\right) \vert 0 \rangle, \, \label{H0}
\ee
where repeated holomorphic and anti-holomorphic indices are summed.
After a little algebraic manipulation it is easy to see that \eqref{HD3+}, \eqref{HD3-} and \eqref{HD30} carries a SU(2) representation in terms of ladder $\pm$ and uncharged t'Hooft symbols (see appendix \ref{appB} for conventions and defintions)
\be
\HH_{1}^{\left( +\right) D3} = -\dfrac{i}{4} \, \eta_{-}^{\mu  \nu } \, T_{\mu  \nu} \, \psi_{1\, 2}  \vert 0 \rangle, \,
\ee
\be
\HH_{1}^{\left( -\right) D3} = +\dfrac{i}{4} \, \eta_{+}^{\mu \nu } \, T_{\mu  \nu} \, \psi_{\bar{1}\, \bar{2}}  \vert 0 \rangle, \,
\ee
\be
\HH_{0}^{D3} = -\dfrac{i}{2} \, \eta_{3}^{\mu  \nu } \, T_{\mu  \nu} \, \vert 0 \rangle, \,
\ee
where we have defined
\be
\eta_{+}^{\mu \nu } \equiv \eta_1^{\mu \nu} +i \eta_2^{\mu \nu} \qquad , \qquad  \eta_{-}^{\mu \nu } \equiv \eta_1^{\mu \nu} -i \eta_2^{\mu \nu}.
\ee
The covariant tensor  $T^{\mu  \nu}$ is given by
\be
T^{\mu  \nu} = \left(\begin{matrix}
 \left[  A^{\mu} ,A^{\nu} \right]  + \dfrac{1}{2} \, \omega_{\alpha} \left( \gamma^{ \mu \nu}\right) ^{\alpha \beta} \bar{\omega}_{\beta} & 0 \\
0 & \left[  a^{\mu} ,a^{\nu} \right]  - \dfrac{1}{2} \, \bar{\omega}_{\alpha} \left( \gamma^{ \mu \nu}\right) ^{\alpha \beta} \omega_{\beta} 
\end{matrix}\right).
\ee

Restricting our attention to  the 4 dimensional worldvolume of the D3-branes, we readily get the quartic potential 
\be
S_T = \tr\left[\bra{\HH_{1}^{(+)D3}}\,\HH_1^{(-)D3}\rangle+\frac14\bra{\HH_{0}^{D3}}\,\HH_0^{D3}\rangle \right]= -\dfrac{1}{16}\, \tr \left[D_a D_a\right] ,
\ee
where 
\be
D_a=\eta_a^{\mu\nu}\, T_{\mu\nu},
\ee
contains (on the $D(-1)$ slot) the ADHM constraints.

For completeness,  the full quartic potential in covariant form is easily computed to be
\be
S_{\rm eff}^{(4)}(\Phi) &=& -\dfrac{1}{8} \, \tr \left[ \left[ A_\mu , A_\nu \right] \left[ A^\mu , A^\nu \right] + \left[ a_\mu ,  a_\nu \right] \left[ a^\mu , a^\nu \right] +  \left[ \phi_i , \phi_j \right] \left[ \phi^m , \phi^j \right] + \left[ \chi_i ,  \chi_j \right] \left[ \chi^m , \chi^j \right] \right] \0\\
&& - \dfrac{1}{4}  \, \tr \left[ \left[A_{\mu}, \phi_i \right] \left[A^{\mu}, \phi^m \right] + \left[a_{\mu}, \chi_i \right] \left[a^{\mu}, \chi^m \right] + \dfrac{1}{4} \left( \omega \gamma^{\mu \nu} \bar{\omega} \right)^2 + \dfrac{1}{4} \left( \bar{\omega} \gamma^{\mu \nu} \omega \right)^2 \right] \0\\
&& - \dfrac{1}{4} \, \tr \left[ \left[A_\mu, A_\nu \right] \omega_\alpha \gamma^{\mu \nu} \bar{\omega}_\beta - \left[a_\mu, a_\nu \right] \bar{\omega}_\alpha \gamma^{\mu \nu} \omega_\beta \right] + \frac{1}{2} \tr \left[ \phi^2 \omega_\alpha \epsilon^{\alpha \beta} \bar{\omega}_\beta - \chi^2 \bar{\omega}_\alpha \epsilon^{\alpha \beta} \omega_\beta \right] \0\\
&& - \tr \left[ \phi_i \, \omega_\alpha \, \chi^m \, \bar{\omega}_\beta \epsilon^{\alpha \beta} \right].
\label{EFF}\ee
If we switch off the D3 degrees of freedom,  this  result is in agreement with  \cite{torinesi} once that a suitable rescaling of the fields is carried out. We also get the algebraic couplings on the $D3$ which were not considered in \cite{torinesi} because they are $\alpha'$ suppressed wrt the couplings on the $D(-1)$'s, by dimensional analysis. It is interesting to note that the coupling $A \, \omega \, a \, \bar{\omega}$ (although  not forbidden in principle) does not appear in the quartic potential \eqref{EFF} since there is not the corresponding auxiliary field.

\section{Conclusions}\label{concl}

In this paper  we have found  that there is an hidden localization mechanism at work on the worldsheet which accounts for the algebraic part of the D-branes massless effective action. These boundary contributions emerge in the process of picture changing, by taking into account that the propagator fails to truly invert the BRST charge. This localization mechanism  can be considered as a rigorous justification to the use of the auxiliary-fields  in the computation of certain superstring amplitudes \cite{torinesi, narain}, in a genuine  zero-momentum setting.

 It would be interesting to extend our analysis to higher orders $O(g^{k>4})$ and to analyze the pattern of the involved, possibly new,  auxiliary-fields. It would be also interesting to see if some couplings involving  space-time fermions localize. These analysis could give new insights on the problem of $\alpha'$ corrections in non-abelian D-brane systems.

It should be noted that the same kind of localization at the boundary of moduli space that we discuss in this paper is at work in the topological string via the holomorphic-anomaly \cite{holom, BCOV} and it would be interesting to explore the  connections.  

From the string field theory perspective we would like to understand if there is an analogous (perhaps more closely related to \cite{Sen-restoration}) localization mechanism in the $A_\infty$ formulation in the small Hilbert space \cite{EKS}. This in particular  could be useful for analyzing loop contributions, after having introduced the Ramond sector \cite{Konopka:2016grr, Erler:2016ybs}. 

We hope that our observations  could be a useful step to better understand the structure of string field theories and how they relate to the low energy effective world.

\section*{Acknowledgments}

 We thank Marco Bill\'o, Ted Erler, Marialuisa Frau, Alberto Lerda,  Yuji Okawa, Igor Pesando,   Ivo Sachs and Martin Schnabl for useful conversations. We are indebted with Ashoke Sen for illuminating discussions about \cite{Sen-restoration}.
 
CM thanks Michael Kroyter and the organizers of the workshop SFT@HIT in Holon, June 2017 where a preliminary version of our result was presented. 
CM thanks  CEICO centre for Theretical Physics at the ASCR, Prague for invitation and for providing a stimulating environment where part of this research was performed.
This work is partially supported by the Compagnia di San Paolo contract {\it MAST: Modern Applications of String Theory} TO-Call3-2012-0088 and by the MIUR PRIN Contract 2015MP2CX4 {\it Non-perturbative Aspects Of Gauge Theories And Strings}.

\appendix 
\section{Details of the localization of the action}\label{appA}

In this section we show the details on the localization of the effective action at quartic order. We consider the propagator term in \eqref{Sprop}:
\be
S^{(4)}_{\rm prop}&=&\dfrac{1}{8}\Tr\left[[\eta_0 \Phi_A,Q_B\Phi_A]\, \xi_0 \dfrac{b_0}{L_0}\bar{P}_0 \, [\eta_0 \Phi_A,Q_B\Phi_A]\right].
\ee
When the decomposition in charged fields of the $\mathcal{N}=2$ is possible, we observe that some contribution appearing in \eqref{Sprop} cannot simultaneously conserve both charge and ghost number. This happens because terms with a propagator force us to pick up the matter vertex operator of weight one $\VV_1$ in $Q_B \Phi_A$ \eqref{stringfieldQ}. For this reason, out of $16$ terms we have at the beginning, only $8$ survive. These  in turn are equal two by two thanks to the properties of the Witten trace. Then we remain with
\be
S^{(4)}_{\rm prop}&=& + \dfrac{1}{4}\Tr\left[[\eta_0 \Phi_A^{(+)},Q_B\Phi_A^{(+)}]\, \xi_0 \dfrac{b_0}{L_0}\bar{P}_0 \, [\eta_0 \Phi_A^{(-)},Q_B\Phi_A^{(+)}]\right] \0\\
&& + \dfrac{1}{4}\Tr\left[[\eta_0 \Phi_A^{(-)},Q_B\Phi_A^{(-)}]\, \xi_0 \dfrac{b_0}{L_0}\bar{P}_0 \, [\eta_0 \Phi_A^{(+)},Q_B\Phi_A^{(-)}]\right]\0\\
&& + \dfrac{1}{4}\Tr\left[[\eta_0 \Phi_A^{(+)},Q_B\Phi_A^{(+)}]\, \xi_0 \dfrac{b_0}{L_0}\bar{P}_0 \, [\eta_0 \Phi_A^{(-)},Q_B\Phi_A^{(-)}]\right]\0\\
&& + \dfrac{1}{4}\Tr\left[[\eta_0 \Phi_A^{(-)},Q_B\Phi_A^{(+)}]\, \xi_0 \dfrac{b_0}{L_0}\bar{P}_0 \, [\eta_0 \Phi_A^{(+)},Q_B\Phi_A^{(-)}]\right],
\label{sprop2}\ee
where the last two terms are symmetric and the first two terms are exchanged if $\Phi_A^{(+)} \leftrightarrow \Phi_A^{(-)}$. Although the first two terms are not zero simply by charge/ghost number conservation it is possible to show that they vanish identically. Since they are symmetric in the exchange $\Phi_A^{(+)} \leftrightarrow \Phi_A^{(-)}$ we show explicitly only that the first term in \eqref{sprop2}       is zero. Due to the on-shell condition on $\Phi_A^{(\pm)}$ we can extract from the commutators $\eta_0$ on the right and $Q_B$ on the left to obtain: 
\be
\Tr\left[\left(Q_B[\Phi_A^{(+)},\eta_0 \Phi_A^{(+)}]\right)\, \xi_0 \dfrac{b_0}{L_0}\bar{P}_0 \, \left(\eta_0 [\Phi_A^{(-)},Q_B\Phi_A^{(+)}]\right)\right].
\ee
Move $\eta_0$ and $Q_B$ in the trace let us write
\be
- \Tr\left[[\Phi_A^{(+)},\eta_0 \Phi_A^{(+)}]\,  \left( \bar{P}_0 - \dfrac{b_0}{L_0}\bar{P}_0  Q_B \right) \,[\Phi_A^{(-)},Q_B\Phi_A^{(+)}]\right].
\ee
The term involving the projector is zero for charge conservation while the other involving the propagator is zero for charge/ghost number conservation. Analogous computations lead the cancellation of the second term in \eqref{sprop2}. So we have to analyze only the two last lines of \eqref{sprop2}:
\begin{itemize}
\item  In the third line we extract from the commutators in a symmetric way $\eta_0$ and $Q_B$, repeating the same computation carried out for the first line to simplify the propagator. This time it is not zero because charge is conserved in terms with the projector $\bar{P}_0$.
\item In the fourth line we use the following identity
\be
[\eta_0 \Phi_A^{(\pm)},Q_B\Phi_A^{(\mp)}] = \eta_0 Q_B [\Phi_A^{(\pm)},\Phi_A^{(\mp)}] + [Q_B \Phi_A^{(\pm)},\eta_0 \Phi_A^{(\mp)}] ,
\ee
on the right and on the left of the propagator. Charge conservation implies the propagator terms cancel out while the remaining ones have a projector inside.
\end{itemize}
These algebraic manipulations allow us to rewrite equation \eqref{sprop2} as the sum of the completely localized effective action we have in (3.17) and some spurious contact terms
\be
S^{(4)}_{\rm prop}&=& S^{(4)}_{\rm eff} + S^{(4)}_{\rm prop,c}\ ,
\ee
where $S^{(4)}_{\rm prop,c}$ collects all the spurious contact terms and is  given by
\be
S^{(4)}_{\rm prop,c} &=& + \dfrac{1}{8}\Tr\left[[\eta_0 \Phi_A^{(-)}, \Phi_A^{(-)}]\,  [\Phi_A^{(+)},Q_B\Phi_A^{(+)}]\right]
+ \dfrac{1}{8}\Tr\left[[\eta_0 \Phi_A^{(+)}, \Phi_A^{(+)}]\,  [\Phi_A^{(-)},Q_B\Phi_A^{(-)}]\right] \0\\
&& + \dfrac{1}{8}\Tr\left[[\Phi_A^{(+)}, \Phi_A^{(-)}]\,  [ \eta_0 \Phi_A^{(+)},Q_B\Phi_A^{(-)}]\right]
+ \dfrac{1}{8}\Tr\left[[\Phi_A^{(-)}, \Phi_A^{(+)}]\,  [ \eta_0 \Phi_A^{(-)},Q_B\Phi_A^{(+)}]\right]. \0\\
\ee
Now we show that these extra contact terms cancel exactly with the contact term coming from the Berkovits action:
\be
S_c^{(4)} = -\dfrac{1}{24}\Tr[[\eta_0 \Phi_A, \Phi_A] \,[\Phi_A, Q_B \Phi_A]].
\ee
Once that the string field $\Phi_A$ is splitted in two charged string fields $\Phi_A^{(\pm)}$, the charge conservation requires that only terms with 2 $\Phi_A^{(+)}$ and 2 $\Phi_A^{(-)}$ survive since the uncharged matter vertex $\VV_1$ in $Q_B \Phi_A$ cannot give contribution. So the contact term is given by the sum of 6 terms:
\be
S_{\rm c}^{(4)} &=& -\dfrac{1}{24}  \Tr \left[ [\eta_0 \Phi_A^{(-)}, \Phi_A^{(-)}] \,[\Phi_A^{(+)}, Q_B \Phi_A^{(+)}]\right] -\dfrac{1}{24} \Tr \left[ [\eta_0 \Phi_A^{(+)}, \Phi_A^{(+)}] \,[\Phi_A^{(-)}, Q_B \Phi_A^{(-)}]\right] \0\\
&& -\dfrac{1}{24} \Tr \left[ [\eta_0 \Phi_A^{(+)}, \Phi_A^{(-)}] \,[\Phi_A^{(+)}, Q_B \Phi_A^{(-)}]\right] -\dfrac{1}{24} \Tr \left[ [\eta_0 \Phi_A^{(+)}, \Phi_A^{(-)}] \,[\Phi_A^{(-)}, Q_B \Phi_A^{(+)}]\right] \0\\
&& -\dfrac{1}{24} \Tr \left[ [\eta_0 \Phi_A^{(-)}, \Phi_A^{(+)}] \,[\Phi_A^{(+)}, Q_B \Phi_A^{(-)}]\right]-\dfrac{1}{24} \Tr \left[ [\eta_0 \Phi_A^{(-)}, \Phi_A^{(+)}] \,[\Phi_A^{(-)}, Q_B \Phi_A^{(+)}]\right]. \0\\
\ee
In order to show that $S_{\rm c}^{(4)} + S^{(4)}_{\rm prop,c} = 0$ we have to make some ordinary but lenghty algebraic manipulations. These are carried out using the properties of the derivations $Q_B$ and $\eta_0$ inside the Witten trace and the Grassmann-graded Jacobi identity
\be
(-1)^{AC} [ \Phi_A , [ \Phi_B, \Phi_C]] + (-1)^{AB} [ \Phi_B , [ \Phi_C, \Phi_A]] + (-1)^{BC} [ \Phi_C , [ \Phi_A, \Phi_B]]=0.
\ee
Three universal structures appear in these computations, we refer to them simply by $C_1, C_2,C_3$ defined as follows:
\be
C_1 =  \dfrac{1}{24} \Tr\left[[\Phi_A^{(-)}, \Phi_A^{(+)}]\,  [ \eta_0 \Phi_A^{(+)},Q_B\Phi_A^{(-)}]\right],
\ee
\be
C_2 = \dfrac{1}{24}\Tr\left[[\Phi_A^{(-)}, \Phi_A^{(+)}]\,  [ \eta_0 \Phi_A^{(-)},Q_B\Phi_A^{(+)}]\right],
\ee
\be
C_3 = -\dfrac{1}{24} \Tr \left[ [\eta_0 \Phi_A^{(+)}, \Phi_A^{(-)}] \,[\Phi_A^{(+)}, Q_B \Phi_A^{(-)}]\right].
\ee
So it is easy to check that every term in $S^{(4)}_{\rm c}$ and $S^{(4)}_{\rm prop,c}$ can be expressed as a linear combination of $C_i$. We report here the final results:
\be
-\dfrac{1}{24} \Tr \left[ [\eta_0 \Phi_A^{(+)}, \Phi_A^{(-)}] \,[\Phi_A^{(-)}, Q_B \Phi_A^{(+)}]\right] = C_1 + C_3,
\ee
\be
-\dfrac{1}{24} \Tr \left[ [\eta_0 \Phi_A^{(-)}, \Phi_A^{(+)}] \,[\Phi_A^{(+)}, Q_B \Phi_A^{(-)}]\right] = C_1 + C_3,
\ee
\be
-\dfrac{1}{24} \Tr \left[ [\eta_0 \Phi_A^{(+)}, \Phi_A^{(+)}] \,[\Phi_A^{(-)}, Q_B \Phi_A^{(-)}]\right] = - C_1 + C_3,
\ee
\be
-\dfrac{1}{24} \Tr \left[ [\eta_0 \Phi_A^{(-)}, \Phi_A^{(+)}] \,[\Phi_A^{(-)}, Q_B \Phi_A^{(+)}]\right] =  C_1 + C_2 + C_3,
\ee
\be
-\dfrac{1}{24} \Tr \left[ [\eta_0 \Phi_A^{(-)}, \Phi_A^{(-)}] \,[\Phi_A^{(+)}, Q_B \Phi_A^{(+)}]\right] =  C_1 + 2 \,C_2 + C_3.
\ee
Then summing all the terms appearing in $S^{(4)}_{\rm c}$ and $S^{(4)}_{\rm prop,c}$ we obtain
\be
S^{(4)}_{\rm c} = 3 \,C_1 + 3 \,C_2 + 6 \,C_3,
\ee
\be
S^{(4)}_{\rm prop,c} = - \left(3 \,C_1 + 3 \,C_2 + 6 \, C_3 \right),
\ee
so that the cancellation of contact terms is demonstrated. This concludes the proof of \eqref{dff}

\section{Supersymmetry, bosonization and correlation functions}\label{appB}

\paragraph{Supersymmetry}
Here we recollect our conventions on supersymmetry notation and the main properties of the t'Hooft symbols. The Euclidean Lorentz group $SO(4)$ on the D3 branes system is realized on spinors in terms of the Pauli matrices $\tau^a$ 
\be
\tau^1 = \left(\begin{matrix} 0 & 1\\
1 & 0 \end{matrix}\right)  \qquad , \qquad \tau^2 = \left(\begin{matrix} 0 & -i\\
i & 0 \end{matrix}\right) \qquad , \qquad \tau^3 = \left(\begin{matrix} 1 & 0\\
0 & -1 \end{matrix}\right),
\ee
from which it is possible to construct the ordinary gamma matrices satisfying the euclidean Clifford algebra. It is well known that the self-dual and antiself-dual generators of SO(4) are two indices gamma matrices $(\gamma^{\mu \nu})_{\alpha} ^{\, \, \, \beta}$, $ (\bar{\gamma}^{\mu \nu})^{ \dot{\alpha }}_{\, \, \, \dot{\beta}} $ given in terms of the self-dual and antiself-dual t'Hooft symbol which maps the Wick rotated Lorentz group to SU(2)
\be
\left( \gamma^{\mu \nu}\right)_{\alpha} ^{\, \, \, \beta} := i \, \eta_c^{\mu \nu} \left( \tau^c \right)_{\alpha} ^{\, \, \, \beta}
\qquad, \qquad \left( \bar{\gamma}^{\mu \nu}\right)^{ \dot{\alpha }}_{\, \, \, \dot{\beta}} := i \, \bar{\eta}_c^{\mu \nu} \left( \tau^c \right)^{ \dot{\alpha }}_{\, \, \, \dot{\beta}}.
\ee
They are symmetric in the spinor indices and antisymmetric in the spacetime indices. In this paper, since we are dealing with instantons we are using only the self-dual two-indices gamma matrices. Self-dual t'Hooft symbols are realized as
\be
\eta_{a \mu \nu} := \epsilon_{a \mu \nu} + \delta_{a \mu} \delta_{4 \nu} - \delta_{a \nu} \delta_{4 \mu},
\ee
and satisfy a a set of properties  including
\be
\eta_{a \mu \nu} = \dfrac{1}{2} \epsilon_{\mu \nu \rho \sigma} \eta_{a \rho \sigma} \qquad , \qquad \eta_{a \mu \nu} \, \eta_{a \rho \sigma} = \delta_{\mu \rho} \delta_{\nu \sigma} - \delta_{\mu \sigma} \delta_{\nu \rho} + \epsilon_{\mu \nu \rho \sigma},
\ee
which are useful for deriving our results in the main text.

Spinor indices are raised and lowered as follows
\be
\psi^{\alpha} = \epsilon^{\alpha \beta} \, \psi_{\beta} \qquad \qquad , \qquad \qquad \psi_{\dot{\alpha}} = \epsilon_{\dot{\alpha}\dot{ \beta}} \, \psi^{\dot{\beta}},
\ee
\be
\psi_{\alpha} = \psi^{\beta} \, \epsilon_{\beta \alpha}  \qquad \qquad , \qquad \qquad \psi^{\dot{\alpha}} =  \psi_{\dot{\beta}} \, \epsilon^{\dot{ \beta}\dot{\alpha}},
\ee
where the $\epsilon$ matrices are defined such that $\epsilon^{12} =  \epsilon_{12} =   \epsilon^{\dot{2}\dot{1}}= \epsilon_{\dot{2}\dot{1}} = 1 $  which implies 
\be
\epsilon^{\beta \alpha} \, \epsilon_{\alpha \gamma} = - \delta_\gamma^\beta \qquad \qquad, \qquad \qquad \epsilon^{\dot{\beta}\dot{\alpha}} \, \epsilon_{\dot{\alpha} \dot{\gamma}} =  - \delta_{\dot{\gamma}}^{\dot{\beta}}.
\ee

\paragraph{Bosonization}

Bosonization in ten dimensions allows us to introduce five commuting scalars $h_i$, one for each complex dimension, and write all the content of the $\psi , S^\alpha$ matter sector in terms of these scalars. Associated to each scalar we have a current proportional to $\partial h_i$ such that
\be
J_i (z) \, e^{\pm i k h_j} (w) \sim \pm k \, \dfrac{\delta_{ij}}{z-w} \,e^{\pm i k h_j} (w).
\ee
We can extend this current to all the spacetime with linear combinations of the five elementary currents. The existence of this current is fundamental to localize the effective action. We have equivalent choices in the full spacetime (for example when we analyze Yang-Mills on a $D(2n)$ branes system as in section 4.1), but when we introduce spin fields on the worldvolume of the D3 branes system we have two inequivalent choices for the total current
\be
J_+ = J_1 + J_2 \qquad  or  \qquad  J_- = J_1 - J_2,
\ee
under which matter fields like $\psi$ are generically charged up to a sign, while spin fields are divided in two families of opposite chiralities
\be
S^{\alpha} = e^{\pm \frac{i}{2}\left(h_1 +h_2 \right)} \qquad , \qquad S^{\dot{\alpha}} = e^{\pm \frac{i}{2}\left(h_1 - h_2 \right)}.
\ee
The bosonized spin fields are charged under one current and uncharged under the other one, so that the choice of the chirality of the spin fields (and hence the choice between $D(-1)$'s or anti $D(-1)$'s) implies the choice of the localizing charge.\\

The bosonized spin fields makes explicit the subleading contribution from the OPEs which are required for the auxiliary fields \eqref{HD3+}, \eqref{HD3-}, \eqref{HD30}:
\be
S^{(\frac12,\frac12)} (z) S^{(\frac12,\frac12)}(w) \sim (z-w)^{\frac{1}{2}} \, e^{i \left(h_1 + h_2 \right)}  = (z-w)^{\frac{1}{2}} \, \psi_{1}\psi_{2} (w),
\ee
\be
S^{(-\frac12,-\frac12)} (z) S^{(-\frac12,-\frac12)}(w) \sim (z-w)^{\frac{1}{2}} \, e^{-i \left(h_1 + h_2 \right)}  = (z-w)^{\frac{1}{2}} \, \psi_{\bar{1} } \psi_{\bar{2} }(w),
\ee
\be
S^{(\frac12,\frac12)} (z) S^{(-\frac12,-\frac12)}(w)\sim (z-w)^{-\frac{1}{2}}.
\ee

\paragraph{Other useful formulas}
Other correlators on the UHP which we use are
\be
\left\langle \psi^{\mu} (z) \psi^{\nu}(w) \right\rangle = \dfrac{\eta^{\mu \nu}}{(z-w)} \quad , \quad
\left\langle c(z_1) c(z_2) c(z_3) \right\rangle = (z_1-z_2)(z_1-z_3)(z_2-z_3) 
\ee
\begin{equation}
\left\langle e^{-\phi} (z) e^{-\phi} (w)\right\rangle = \dfrac{1}{z-w} \qquad, \qquad 
\left\langle \xi (z) \eta (w) \right\rangle = \dfrac{1}{z-w},
\end{equation}
and correlation functions involving more than two of the above fields are easily derived by Wick theorem. The twist field two-point  functions are given as  in  \cite{torinesi}
\begin{equation}
\left\langle \Delta(z) \bar{\Delta}(w) \right\rangle = (z-w)^{-\frac{1}{2}}  \qquad , \qquad \left\langle \bar{\Delta}(z) \Delta(w) \right\rangle = -(z-w)^{-\frac{1}{2}},
\end{equation}
where an effective minus sign is present in the second two-point function to account for the correct odd grassmanality of the superconformal primary $\Delta S^\alpha$.  We refer to \cite{torinesi} for more details. In this work we  also use the two following OPE's 
\be
S^{\alpha} (z) \, S^{\beta} (w) \, &\sim& \,  \dfrac{\epsilon^{\alpha \beta}}{(z-w)^{\frac{1}{2}}} -\frac14 (z-w)^{\frac12}(\gamma_{\mu\nu})^{\alpha\beta}\psi^{\mu}\psi^{\nu},\label{pop}\\
\psi^\mu (z) \, S^{\alpha} (w) \, &\sim& \, - \dfrac{1}{\sqrt{2}} \, \dfrac{\left( \gamma^{\mu} \right)^{ \alpha}_{ \, \, \, \,  \,  \dot{\beta}} \, S^{\dot{\beta}} (w)}{(z-w)^{\frac{1}{2}}}. \label{pip}
\ee
Notice that the first term in \eqref{pop},  proportinal to the identity, is responsible for the auxiliary field $\HH_0$, while the second term, which is subleading, is responsible for the charged auxiliary fields $\HH_1^{\pm}$.   The other OPE \eqref{pip}  is responsible for the absence of an auxiliary field giving rise to an $Awa\bar w$ coupling, since when, properly dressed with a twist field and the ghosts, the weight zero contribution is not present.

\end{document}